\documentclass[preprintnumbers,superscriptaddress,showpacs,pra]{revtex4}%
\usepackage[colorlinks=true,linkcolor=blue]{hyperref}
\usepackage{amssymb}
\usepackage{amsfonts}
\usepackage{amsmath}
\usepackage{graphicx}%
\providecommand{\U}[1]{\protect\rule{.1in}{.1in}}
\let\stdsection\section
\renewcommand\section{\nopagebreak\stdsection}
\begin{document}
\title{Geometric Momentum as a Probe of Embedding Effect}
\author{Q. H. Liu}
\affiliation{School for Theoretical Physics, and Department of Applied Physics, Hunan
University, Changsha, 410082, China. Tel/fax 86-731-88820378/86-731-88822332,
Email: quanhuiliu@gmail.com}

\begin{abstract}
As a submanifold is embedded into higher dimensional flat space, quantum
mechanics gives various embedding quantities. In the present study, two
embedding quantities for a two-dimensional curved surface examined in
three-dimensional flat space, the geometric momentum and the geometric
potential, are derived in a unified manner. Then for a particle moving on a
two-dimensional sphere or a free rotation of a spherical top, the projections
of the geometric momentum $\mathbf{p}$ and the angular momentum $\mathbf{L}$
onto a certain Cartesian axis form a complete set of commuting observables as
$[p_{i},L_{i}]=0$ ($i=1,2,3$), thus constituting a dynamical ($p_{i},L_{i}$)
representation for the states on the two-dimensional spherical surface. The
geometric momentum distribution of the state represented by spherical
harmonics is successfully obtained, and this distribution for a homonuclear
diatomic molecule seems within the resolution power of present momentum
spectrometer and can be measured to probe the embedding effect.

\end{abstract}
\date{\today}

\maketitle
\preprint{REV\TeX4-1}

\section{Introduction}

In microscopic domain, there are many quantum motions confined on the
two-dimensional surfaces, e.g., mobile carriers on the corrugating graphene
sheet or spherical fullerene molecule $C_{60}$, and free rotation of hydrogen
nuclei around its center of mass in a hydrogen molecule at relatively low
temperature, etc. Over the past decade, we have witnessed that physics
community gradually reaches consensus that the examination of the quantum
motions confined on the two-dimensional surfaces in three-dimensional
Euclidean space $R^{3}$ is of physical significance. In the Euclidean space,
the quantum physics is well defined, one can use conventional quantum
mechanics without any new postulate imposed \cite{dirac1,dirac2}. The
confinement of the particle on a curved two-dimensional manifold is treated as
the limiting case of a particle in a three-dimensional manifold in which
\textit{a confining potential} is invoked to acting in the normal direction of
its two-dimensional surface, and we can then have an unambiguous formulation
of quantum mechanics on the surface as a secondary or derived theory
\cite{jk,dacosta,1992,kleinert,1997-1,1998,FC,JD,OB,liu11-1,SOint1,SOint2,SOint3,SOint4,2010-2,epl}%
. However, if one hopes to reproduce the same theory within the Dirac's theory
for a systems with second class constraints \cite{dirac2}, some cautions need
to be taken into consideration \cite{note0}. The new formulation of quantum
mechanics on the surface involves both gaussian and mean curvature. In simple
and plain words, the mean curvature $M$, on one hand, is an extrinsic
curvature that is not detectable to someone who can not study the
three-dimensional space surrounding the surface on which he resides, whereas
the gaussian curvature $K$, on the other hand, is an intrinsic curvature that
is detectable to the \textquotedblleft two-dimensional
inhabitants\textquotedblright\ on a surface and not just outside observers
\cite{wolfram}. In purely intrinsic geometry, undefinable and even meaningless
is the shape itself of a surface.

When no electromagnetic field is applied and the spin of the particle plays
insignificant role, the marked feature of the theory is the dependence of both
an effective potential $V_{g}$ \cite{note1} in the Hamiltonian and the
geometric momentum $\mathbf{p}$ on the mean curvature $M$
\cite{jk,dacosta,FC,JD,OB,liu11-1}. The geometric potential
\cite{jk,dacosta,note1} with $\mu$ denoting the mass,%
\begin{equation}
V_{g}=-\hbar^{2}/(2\mu)(M^{2}-K) \label{1}%
\end{equation}
comes from how to define a proper form of Laplacian operator acting on a
quantum state on surface \cite{liu11-1,2007}, whereas the geometric momentum,
with $\mathbf{\nabla}_{2}$ being the gradient operator on a two-dimensional
surface \cite{diffgeom} and $\mathbf{n}$ standing for the normal vector of the
surface at a given point,
\begin{equation}
\mathbf{p}=-i\hbar(\mathbf{\nabla}_{2}+M\mathbf{n}), \label{vggm}%
\end{equation}
is related to a proper form of gradient operator on the state \cite{liu11-1}.
When first exposed to this expression (\ref{vggm}) apparently containing a
term $M\mathbf{n}$, many thinks it has component along the normal direction,
but it is not the case. Actually, it is an operator exclusively defined on the
tangent plane to surface at the given point for we have an operator relation
$\mathbf{p}\cdot\mathbf{n+n}\cdot\mathbf{p}=0$ with use of a relation$\ \nabla
_{2}\cdot\mathbf{n}=-2M$ \cite{diffgeom}. One can also call (\ref{1}) and
(\ref{vggm}) the embedding potential \cite{japan1993} and the embedding
momentum, respectively. As we see in the appendix, both geometric quantities
(\ref{1}) and (\ref{vggm}) can be derived within the same theoretical
framework. An experimental verification of the potential amounts to an
indirect affirmative experimental evidence of the momentum as well, and
\textit{vice versa}. The geometric potential has recently been experimentally
verified \cite{2010-2,epl}, and it is an important advance in quantum
mechanics, implying that quantum mechanics based on purely intrinsic geometry
does not offer a proper description of the constrained motions in microscopic
domain, provided that the extrinsic examination is performed as well. Here we
mention that the spin of the particle usually plays a role via the surface
spin-orbit coupling \cite{SOint1,SOint2,SOint3}, etc.\cite{SOint4}, obtained
also from the same procedure of squeezed limit of its the three-dimensional analogue.

Noting that the linear momentum distribution of an electron state within a
hydrogen atom can be easily carried out and had been experimentally verified
\cite{MomentumSpect1,MomentumSpect2}. Let us consider the simplest constrained
motion on two dimensional spherical surface $S^{2}$ and ask whether it is
possible to give a momentum space representation for the states on it. An
immediate problem is what the\ proper momentum is. It can never be the usual
linear momentum $-i\hbar\mathbf{\nabla=}-i\hbar\mathbf{(\partial}%
_{x},\mathbf{\partial}_{y},\mathbf{\partial}_{z}\mathbf{)}$ because the motion
on $S^{2}$ has only two degrees of freedom while $-i\hbar\mathbf{\nabla}$ has
three mutually commutable components that are too many to form a complete set
of commuting observables for $S^{2}$. Moreover, as we stress before
\cite{liu11-1}, a set of self-adjoint momentum operators in purely intrinsic
geometry is unattainable for any states on $S^{2}$. In addition, the geometric
momentum $\mathbf{p}=-i\hbar(\mathbf{\nabla}_{2}+M\mathbf{n})$ (\ref{vggm})
alone does not suffice because its three components are not mutually
commutable, thus too few to provide a complete set of commuting observables.
The key finding of the present study offers a solution to the problem, based
on a discovery of a new dynamical representation on the surface.

The organization of the present paper is as follows. In next sections II, we
present a unified derivation of both geometric potential (\ref{1}) and
momentum (\ref{vggm}). In next sections III, we starts from a dynamical
symmetric group $SO(3,1)$ on the sphere which yields a proper and complete set
of commuting observables, to arrive at a dynamical representation mixing the
geometric momentum and orbital angular momentum. In the section IV, it aims at
the explicit form of the geometric momentum distribution of the some molecular
rotational states. Section V gives a brief discussion of the results obtained,
and conclude the present study.

\section{Geometric momentum for a particle on a curved surface}

To get the geometric momentum (\ref{vggm}), we utilize exactly the same manner
how the geometric potential is derived \cite{jk,dacosta,FC}. For ease of the
comparison, we use similar set of symbols as Ferrari and Cuoghi who recently
build up a theoretical framework with geometric potential when the
electromagnetic field is applied \cite{FC}. The lowercase Latin letters
$i,j,k$ stand for the 3D indices and assume the values $1,2,3$, e.g.,
$\mathbf{(}x_{i},p_{j}\mathbf{)}$ for the position and momentum in 3D
Cartesian coordinates. Position specified by ($q^{1},q^{2},q^{3}$) can be
understood as description of the position in the curvilinear coordinates
parameterizing a manifold. Now let the 2D surface under study is considered as
a more realistic 3D shell whose equal thickness $d$ is negligible in
comparison with the dimension of the whole system. The position $\mathbf{R}$
within the shell in the vicinity of the surface $S$ can be parametrized as
with $0$ $\leq q^{3}\leq d,$
\begin{equation}
\mathbf{R}(q^{1},q^{2},q^{3})=\mathbf{r}(q^{1},q^{2})+q^{3}\mathbf{n}%
(q^{1},q^{2}), \label{3dR}%
\end{equation}
where $\mathbf{r}(q^{1},q^{2})$ parametrizes the surface and $\mathbf{n}%
(q^{1},q^{2})$ denotes the unit normal vector at point $(q^{1},q^{2})$. The
gradient operator $\nabla$ in 3D flat space, expressed in the curvilinear
coordinates, takes following form \cite{diffgeom},
\begin{equation}
\nabla=\mathbf{r}^{\mu}\partial_{\mu}+\mathbf{n}\partial_{q^{3}}\text{,}
\label{grad}%
\end{equation}
where $\mathbf{\nabla}_{2}\equiv\mathbf{r}^{\mu}\partial_{\mu}$ is the
gradient operator on 2D curved surface \cite{diffgeom}. The relation between
the 3D metric tensor $G_{ij}$ and the 2D one $g_{\mu\nu}$ is given by
\cite{dacosta,FC},
\begin{align}
G_{ij}  &  =g_{\mu\nu}+\left[  \alpha g+(\alpha g)^{T}\right]  _{\mu\nu}%
q^{3}+(\alpha g\alpha^{T})_{\mu\nu}\left(  q^{3}\right)  ^{2},\nonumber\\
G_{\mu3}  &  =G_{3\mu}=0,\;G_{33}=1, \label{metric}%
\end{align}
where $\alpha_{\mu\nu}$ is the Weingarten curvature matrix for the surface,
and $M=-\mathrm{Tr}(\alpha)/2$, and $K=\mathrm{\det}(\alpha)$ \cite{dacosta}.
The covariant Schr\"{o}dinger\ equation for particles moving within a thin
shell of thickness $d$ in 3D is \cite{FC}, with presence of both the magnetic
field via the vector potential $\mathbf{A}$ and the electric field via the
scalar potential $V$,
\begin{equation}
\mathrm{i}\hbar\frac{\partial}{\partial t}\psi(\mathbf{q},t)=-\frac{\hbar^{2}%
}{2m}G^{ij}D_{i}D_{j}\psi(\mathbf{q},t)+QV\psi(\mathbf{q},t), \label{schr1}%
\end{equation}
where $Q$ is the charge of the particle and $D_{j}=\nabla_{j}-(\mathrm{i}%
Q/\hbar)A_{j}$ with $A_{j}$ being the covariant components of the vector
potential $\mathbf{A}$. Conveniently denoting the scalar potential $A_{0}=-V$,
we can define a gauge covariant derivative for the time variable as
$D_{0}={\partial_{t}}-{\mathrm{i}Q}A_{0}/\hbar$. The gauge transformations in
quantum mechanics are \cite{FC},
\begin{equation}
A_{j}\rightarrow A_{j}^{\prime}=A_{j}+\partial_{j}\gamma;\text{ }%
A_{0}\rightarrow A_{0}^{\prime}=A_{0}+{\partial_{t}}\gamma;\text{ }%
\psi\rightarrow\psi^{\prime}=\psi\mathrm{e}^{\mathrm{i}Q\gamma/\hbar},
\label{GTran}%
\end{equation}
where $\gamma$ is a scalar function. The Eq.(\ref{schr1}) can be rewritten as
an explicit gauge invariant form \cite{FC},
\begin{equation}
\mathrm{i}\hbar D_{0}\psi=-\frac{\hbar^{2}}{2m}G^{ij}D_{i}D_{j}\psi.
\label{schr2}%
\end{equation}

Now we recall two important facts regarding the wave functions: 1, the
normalization of the wave functions remains whatever coordinates are used, and
the transformation of volume element satisfies $d^{3}\mathbf{x=}\sqrt{G}%
d^{3}\mathbf{q}$ \cite{FC},
\begin{equation}
\int\left\vert \psi(\mathbf{x},t)\right\vert ^{2}\mathrm{d}^{3}\mathbf{x=}%
\int\left\vert \psi(\mathbf{q},t)\right\vert ^{2}\sqrt{G}\mathrm{d}%
^{3}\mathbf{q}=1, \label{Totalnormal}%
\end{equation}
where \cite{dacosta,FC},
\begin{equation}
G=\mathrm{det}(G_{ij})=g\left(  1-2Mq^{3}+K\left(  q^{3}\right)  ^{2}\right)
^{2}.
\end{equation}
2, an advantage of the coordinates (\ref{3dR}) is that the wave function
$\psi(\mathbf{q},t)$ from (\ref{schr1}) or (\ref{schr2}) takes following
factorization form \cite{dacosta,FC},
\begin{equation}
\psi(\mathbf{q},t)=\frac{\chi(q^{1},q^{2},t)}{\sqrt{1-2Mq^{3}+K\left(
q^{3}\right)  ^{2}}}\varphi(q^{3},t),
\end{equation}
and it is guaranteed with suitable choice of gauge for $\gamma$ such that
$A_{3}^{\prime}=0$ \cite{FC},
\begin{equation}
\gamma(q^{1},q^{2},q^{3})=-\int_{0}^{q^{3}}A_{3}(q^{1},q^{2},q)\mathrm{d}q.
\label{ggauge}%
\end{equation}
Combining these two facts, we have two conservations of norm from
(\ref{Totalnormal}),%
\begin{equation}
\oint\left\vert \chi(q^{1},q^{2},t)\right\vert ^{2}\sqrt{g}\mathrm{d}%
q^{1}\mathrm{d}q^{2}=1,\text{ and }\int_{0}^{d}\left\vert \varphi
(q^{3},t)\right\vert ^{2}\mathrm{d}q^{3}=1.
\end{equation}

We are now ready to examine the gradient operator $\nabla$ (\ref{grad}) acting
on the state $\psi(\mathbf{q},t)$ and the result is,%
\begin{align}
\nabla\psi(\mathbf{q},t)  &  =\mathbf{r}^{\mu}\partial_{\mu}\psi
(\mathbf{q},t)+\mathbf{n}\frac{M-q^{3}K}{\left(  1-2Mq^{3}+K\left(
q^{3}\right)  ^{2}\right)  ^{3/2}}\chi(q^{1},q^{2},t)\varphi(q^{3}%
,t)\nonumber\\
&  +\mathbf{n}\frac{\chi(q^{1},q^{2},t)}{\sqrt{1-2Mq^{3}+K\left(
q^{3}\right)  ^{2}}}\partial_{q^{3}}\varphi(q^{3},t).
\end{align}
Then taking limit $d\rightarrow0$, we have $\nabla$ as its acting on the state
$\psi(\mathbf{q},t)$,
\[
\nabla\psi(\mathbf{q},t)=\left(  \mathbf{r}^{\mu}\partial_{\mu}+M\mathbf{n}%
\right)  \psi(\mathbf{q},t)+\mathbf{n}\chi(q^{1},q^{2},t)\partial_{q^{3}%
}\varphi(q^{3},t)
\]
which shows that the gradient operator $\nabla$ can be decomposed into two
separate parts, one part $\left(  \mathbf{r}^{\mu}\partial_{\mu}%
+M\mathbf{n}\right)  $ lies on the tangent plane to surface at a given point
and another is along the direction of normal $\mathbf{n}$, corresponding to
the decomposition of the Schr\"{o}dinger equation into two Schr\"{o}dinger
ones determining $\chi(q^{1},q^{2},t)$ and $\varphi(q^{3},t)$ respectively
\cite{FC}. Paying attention to the motion on the surface only, we have the
resultant operator, $\mathbf{r}^{\mu}\partial_{\mu}+M\mathbf{n}$. With a
coefficient $-i\hbar$ multiplied, the geometric momentum (\ref{vggm}) is derived.

The gauge invariance of the momentum operator $\mathbf{p}=-i\hbar
(\mathbf{r}^{\mu}\partial_{\mu}+M\mathbf{n)-}Q\mathbf{A}$ is assured in the
presence of the vector potential $\mathbf{A}$ with vanishing component along
the normal direction as $A_{3}=0$ being pre-imposed. Under 2D gauge
transformation: $\mathbf{A}\rightarrow\mathbf{A}^{\prime}=\mathbf{A+r}^{\mu
}\partial_{\mu}\gamma$ with $\gamma=\gamma(q^{1},q^{2})$ and $\psi
\rightarrow\psi^{\prime}=\mathrm{e}^{\mathrm{i}Q\gamma/\hbar}\psi$, we have
$\mathbf{p}\psi\rightarrow\mathbf{p}^{\prime}\psi^{\prime}=\mathrm{e}%
^{\mathrm{i}Q\gamma/\hbar}\mathbf{p}\psi$,
\begin{align}
\mathbf{p}^{\prime}\psi^{\prime}  &  =\left(  -i\hbar(\mathbf{r}^{\mu}%
\partial_{\mu}+M\mathbf{n)-}Q(\mathbf{A}+\mathbf{r}^{\mu}\partial_{\mu}%
\gamma)\right)  \psi\mathrm{e}^{\mathrm{i}Q\gamma/\hbar}\nonumber\\
&  =\mathrm{e}^{\mathrm{i}Q\gamma/\hbar}\left(  -i\hbar(\mathbf{r}^{\mu
}\partial_{\mu}+M\mathbf{n)-}Q\mathbf{A}\right)  \psi\nonumber\\
&  =\mathrm{e}^{\mathrm{i}Q\gamma/\hbar}\mathbf{p}\psi.
\end{align}

Noting that there is no direct connection between $\Delta_{LB}+(M^{2}-K)$ and
$\nabla_{2}+M\mathbf{n}$ such as in 3D flat space $\nabla^{2}\equiv\nabla
\cdot\nabla$. For reaching $\Delta_{LB}+(M^{2}-K)$, we have to start from the
Laplace operator in flat 3D space$\ \nabla^{2}=\left(  \mathbf{r}^{\mu
}\partial_{\mu}+\mathbf{n}\partial_{q^{3}}\right)  \cdot\left(  \mathbf{r}%
^{\mu}\partial_{\mu}+\mathbf{n}\partial_{q^{3}}\right)  =\Delta_{LB}%
-2M\partial_{q^{3}}+\partial_{q^{3}}^{2}$ \cite{diffgeom}, then resort to the
confining procedure. Explicitly we have,
\begin{align}
\nabla^{2}\psi(\mathbf{q},t)  &  =\Delta_{LB}\psi(\mathbf{q},t)+\frac
{(M^{2}-K)+2M(2M^{2}-K)q^{3}+2K(K-3M^{2})\left(  q^{3}\right)  ^{2}%
+2MK^{2}\left(  q^{3}\right)  ^{3}}{\left(  1-2Mq^{3}+K\left(  q^{3}\right)
^{2}\right)  ^{5/2}}\psi(\mathbf{q},t)\nonumber\\
&  -\frac{2q^{3}(K-2M^{2}+KMq^{3})}{\left(  1-2Mq^{3}+K\left(  q^{3}\right)
^{2}\right)  ^{3/2}}\chi(q^{1},q^{2},t)\partial_{q^{3}}\varphi(q^{3}%
,t)+\frac{1}{\left(  1-2Mq^{3}+K\left(  q^{3}\right)  ^{2}\right)  ^{1/2}}%
\chi(q^{1},q^{2},t)\partial_{q^{3}}^{2}\varphi(q^{3},t) \label{-2}%
\end{align}
In the same limit limit $d\rightarrow0$, this operator $\nabla^{2}$ (\ref{-2})
becomes,%
\begin{equation}
\nabla^{2}=\Delta_{LB}+(M^{2}-K)+\partial_{q^{3}}^{2} \label{-1}%
\end{equation}
Noting that kinetic energy is $T=-\hbar^{2}/(2\mu)\nabla^{2}$, we see that the
effective potential, the geometric potential as $-\hbar^{2}/(2\mu)(M^{2}-K)$
(\ref{1}) comes out. However, it is a puzzling fact that there is a direct
connection between $\left(  \mathbf{r}^{\mu}\partial_{\mu}+M\mathbf{n}\right)
$ and$\ \Delta_{LB}$ \cite{2007}, as we pointed out in 2007. What puzzling?
the quantity $\left(  \mathbf{r}^{\mu}\partial_{\mu}+M\mathbf{n}\right)  $
involves extrinsic curvature, whereas the quantity $\Delta_{LB}$ comes purely
from intrinsic geometry.

Thus, a unified derivation of both geometric potential and momentum is thus
fulfilled. In the rest part of the present paper, we apply the geometric
momentum (\ref{vggm}) to motion constrained on the two dimensional spherical
surface $S^{2}$.

\section{Geometric momentum -- angular momentum representation on $S^{2}$}

For our purpose to reveal the geometric momentum -- angular momentum
representation on $S^{2}$, we first point out a dynamical $SO(3,1)$ symmetry
on two-dimensional spherical surface, and second present the basic vectors in
($\theta,\varphi$) representation and a new but intermediate representation
respectively, and finally reach the dynamical representation determined by two
mutually commutable quantities.

\subsection{A dynamical $SO(3,1)$ symmetry on two-dimensional spherical
surface}

On the two-dimensional spherical surface of fixed radius $r$ with the mean
curvature $M=-1/r$, three Cartesian components of the geometric momentum
$\mathbf{p}$ are\ from (\ref{vggm}) \cite{liu11-1,2007,2003,liu10,liu13},
\begin{align}
p_{x}  &  =-i\hbar(\cos\theta\cos\varphi\frac{\partial}{\partial\theta}%
-\frac{\sin\varphi}{\sin\theta}\frac{\partial}{\partial\varphi}-\sin\theta
\cos\varphi),\label{hpx}\\
p_{y}  &  =-i\hbar(\cos\theta\sin\varphi\frac{\partial}{\partial\theta}%
+\frac{\cos\varphi}{\sin\theta}\frac{\partial}{\partial\varphi}-\sin\theta
\sin\varphi),\label{hpy}\\
p_{z}  &  =i\hbar(\sin\theta\frac{\partial}{\partial\theta}+\cos\theta),
\label{hpz}%
\end{align}
where the transformation ${p}_{{i}}r\rightarrow p_{i}$ is made to conveniently
convert the momentum into dimension of the angular momentum, i.e., the
dimension of Planck's constant $\hbar$. The three Cartesian components of the
orbital angular momentum $\mathbf{L}$ are well-known as ${L}_{{x}}=i\hbar
(\sin\varphi\partial_{\theta}+\cot\theta\cos\varphi\partial_{\varphi})$,
${L}_{{y}}=-i\hbar(\sin\varphi\partial_{\theta}-\cot\theta\sin\varphi
\partial_{\varphi})$ and ${L}_{{z}}=-i\hbar\partial_{\varphi}$. A derivation
of (\ref{hpx})-(\ref{hpz}) from Dirac's theory is discussed in \cite{liu11-1}
and is commented in \cite{note2}. For a two-dimensional spherical space
$S^{2}$, the constantness of the radius $r$\ is nothing but a parameter
characterizing how curve the space is. For more realistic molecular state such
as homonuclear diatomic molecule, this radius $r$ corresponds to a mean value
$\left\langle r\right\rangle $ whereas $1/r$ corresponds to $\left\langle
1/r\right\rangle $. However, because there is usually no coupling between
radial motion and rotation, the rotational motion can be separated and its
geometric momentum spectrometry can be established individually.

We can easily verify the following commutation relations that form an
$so(3,1)$ algebra \cite{liu11-1}:%
\begin{equation}
\lbrack{p}_{{i}}{,p}_{j}]=-i\hbar\varepsilon_{ijk}L_{k}\text{, }[{L}_{{i}}%
{,p}_{j}]=i\hbar\varepsilon_{ijk}{p}_{k}\text{, }[{L}_{{i}}{,L}_{j}]={i}%
\hbar\varepsilon_{ijk}L_{k}. \label{so31}%
\end{equation}
We see that the quantum motion on the sphere of geometric $O(3)$ symmetry
possesses a dynamical $SO(3,1)$ symmetry. Three commutable pairs (${L}_{{i}%
}{,p}_{i}$) are equivalent with each other upon a rotation of coordinate
system \cite{liu11-1,liu12},
\begin{equation}
f_{x}=\exp(-i\pi L_{y}/2)f_{z}\exp(i\pi L_{y}/2),\text{ }f_{y}=\exp(i\pi
L_{x}/2)f_{z}\exp(-i\pi L_{x}/2),\text{ }(f_{i}\rightarrow L_{i}\text{ or
}p_{i}). \label{rotation}%
\end{equation}
Here we follow the convention that a rotation operation affects a physical
system itself \cite{book6}. Equation (\ref{rotation}) above implies that it is
sufficient to study one representation determined by one pair of the three
$({L}_{{i}}{,p}_{i})$.

\subsection{Eigenfunctions of $(p_{z},L_{z})$ in ($\theta,\varphi$)
representation and a new ($u,\varphi$) representation}

Because motion on $S^{2}$ has two degrees of freedom, a representation needs a
complete set of a complete set of two commuting observables. The well-known
set is the spherical harmonics $Y_{lm}(\theta,\varphi)$\ determined by the
commutable pairs $(L^{2},L_{z})$ in the ($\theta,\varphi$) representation. For
convenience of a comparison between the basis vectors $Y_{lm}(\theta,\varphi)$
and the new ones given by simultaneous functions of both the geometric and the
angular momentum, we choose the $z$-axis component pair $(p_{z},L_{z})$ rather
than $(p_{x},L_{x})$ or $(p_{y},L_{y})$. The common operator $L_{z}$ means
also a choice of the reference direction in position space.

The complete set of the simultaneous eigenfunctions for $(p_{z},L_{z})$ is
given by,%
\begin{equation}
\psi_{p_{z},m}(\theta,\varphi)=\frac{1}{\sqrt{2\pi\hbar}}\frac{1}{\sin\theta
}{\exp\left(  -i\frac{{{p}_{z}}}{\hbar}\ln\tan\frac{\theta}{2}\right)  }%
\frac{1}{\sqrt{2\pi}}e^{im\varphi}. \label{gmlz}%
\end{equation}
The eigenvalues of $(p_{z},L_{z})$ acting on $\psi_{p_{z},m}(\theta,\varphi)$
above are ($p_{z},m\hbar$) respectively. The normalization relation can be
easily verified,
\begin{align}
&  \oint{\psi_{{{{{p}^{\prime}}}_{z}m}^{\prime}}^{\ast}}\left(  \theta
,\phi\right)  {{\psi}_{{{p}_{z},m}}}\left(  \theta,\varphi\right)  \sin\theta
d\theta d\varphi\nonumber\\
&  =\delta_{m^{\prime}m}\frac{1}{2\pi\hbar}\int_{0}^{\pi}{\exp\left(
i\frac{\left(  {{{{p}^{\prime}}}_{z}}-{{p}_{z}}\right)  }{\hbar}(\ln\tan
\frac{\theta}{2})\right)  }\frac{1}{\sin\theta}d\theta\nonumber\\
&  =\delta_{m^{\prime}m}\frac{1}{2\pi\hbar}\int_{-\infty}^{\infty}{\exp\left(
i\frac{\left(  {{{{p}^{\prime}}}_{z}}-{{p}_{z}}\right)  }{\hbar}u\right)
}du\nonumber\\
&  =\delta_{m^{\prime}m}\delta\left(  {{{{p}^{\prime}}}_{z}}-{p}_{{z}}\right)
, \label{normalization}%
\end{align}
where the variable transformation
\begin{equation}
\ln\tan(\theta/2)\rightarrow u\text{, or }\theta\rightarrow2\arctan
(e^{u}),(u\in(-\infty,\infty)), \label{vartransform}%
\end{equation}
is used, and $\delta_{m^{\prime}m}$ is the Kronecker delta that equals to $1$
once $m^{\prime}=m$ and to zero otherwise. This variable transformation
(\ref{vartransform}) has the following profound consequence: It makes the
operator $p_{z}$ (\ref{hpz}) behave like a linear momentum which is defined on
flat space $u\in(-\infty,\infty)$,
\begin{equation}
p_{z}(\theta)\longrightarrow p_{z}(u)=i\hbar\frac{\partial}{\partial u},
\label{momentum}%
\end{equation}
whose eigenfunction is well-known as $exp(-iu{{p}_{z}/}\hbar)/\sqrt{2\pi\hbar
}$ corresponding to eigenvalue ${{p}_{z}}$.

To approach the $(p_{z},L_{z})$ representation of the operators and states, it
is very convenient to utilize the same variable transformation
(\ref{vartransform}) and to use $u$ instead of $\theta$ in all relevant states
and operators. For square of the angular momentum operator,
\begin{equation}
L^{2}(\theta,\varphi)\equiv-\hbar^{2}(\frac{\partial^{2}}{\partial\theta^{2}%
}+\cot\theta\frac{\partial}{\partial\theta}+\frac{1}{\sin^{2}\theta}%
\frac{\partial^{2}}{\partial\varphi^{2}}), \label{l2}%
\end{equation}
we find,%
\begin{equation}
L^{2}(\theta,\varphi)\rightarrow L^{2}(u,\varphi)\equiv-\hbar^{2}\cosh
^{2}({{{{u}}}})\left(  \frac{\partial^{2}}{\partial{{{{u}}}}^{2}}%
+2\tanh({{{{u}}}})\frac{\partial}{\partial{{{{u}}}}}+\frac{\partial^{2}%
}{\partial\varphi^{2}}+1\right)  . \label{l2u}%
\end{equation}
Hereafter, the same operator $L$ with different variables ($\theta,\varphi$)
or ($u,\varphi$) in different representation has a different definition as
clearly shown in (\ref{l2}) and (\ref{l2u}) respectively. By mean of either
directly solving the eigenvalue equation $L^{2}(u,\varphi)Y_{lm}^{\prime
}(u,\varphi)=\lambda Y_{lm}^{\prime}(u,\varphi)$ or by the variable
transformation, the spherical harmonics $Y_{lm}(\theta,\varphi)$ in
($\theta,\varphi$) representation becomes $Y_{lm}^{\prime}(u,\varphi)$ in the
new ($u,\varphi$) representation,
\begin{equation}
Y_{lm}(\theta,\varphi)\rightarrow Y_{lm}^{\prime}(u,\varphi)\equiv N_{lm}%
\frac{P_{l}^{m}(-\tanh u)}{\cosh u}\frac{1}{\sqrt{2\pi}}e^{im\varphi},
\label{Yu}%
\end{equation}
where $m=-l,-l+1,...,-1,0,1,...,l-1,l$, and%
\begin{equation}
N_{lm}=\sqrt{\frac{2l+1}{2}\frac{\left(  l-1\right)  !}{\left(  l+m\right)
!}}.
\end{equation}
The normalization of the spherical harmonics $Y_{lm}^{\prime}(u,\varphi)$
satisfies,
\begin{equation}
\delta_{l^{\prime}l}\delta_{m^{\prime}m}=\oint Y_{l^{\prime}m^{\prime}}^{\ast
}(\theta,\varphi)Y_{lm}(\theta,\varphi)\sin\theta d\theta d\varphi=\int
_{0}^{2\pi}d\varphi\left[  \int_{-\infty}^{\infty}Y_{l^{\prime}m^{\prime}%
}^{\prime\ast}(u,\varphi)Y_{lm}^{\prime}(u,\varphi)du\right]  . \label{norm}%
\end{equation}
It implies that the transformed system is defined on two dimensional stripe
space: $u\in(-\infty,\infty)\cup$ $\varphi\in(0,2\pi)$.

\subsection{States and spherical harmonics in $(p_{z},L_{z})$ representation}

Two operators $(p_{z},L_{z})$ in their own representation is determined by,
\begin{equation}
\hat{p}_{z}\delta(p_{z}-p_{z}^{\prime})=p_{z}^{\prime}\delta(p_{z}%
-p_{z}^{\prime}),\text{ \ }\hat{L}_{z}\delta_{L_{z},m\hbar}=m\hbar
\delta_{L_{z},m\hbar}\text{,} \label{pzlz}%
\end{equation}
where operator $f$ is now denoted with a hat as $\hat{f}$ for avoiding
possible confusion, and symbol $f$ without the hat stands for a variable in
the eigenfunction such as $p_{z}$ in $\delta(p_{z}-p_{z}^{\prime})$ or $L_{z}$
in $\delta_{L_{z},m\hbar}$. In general, a state $\Phi(p_{z},L_{z})$ in
$(p_{z},L_{z})$ representation corresponding to $\Psi(u,\varphi)$ in position
representation is given by,%
\begin{equation}
\Phi(p_{z},L_{z})=\int_{0}^{2\pi}\frac{e^{-im\varphi}}{\sqrt{2\pi}}%
d\varphi\left[  \int_{-\infty}^{\infty}\Psi(u,\varphi)\frac{{\exp\left(
i\frac{{{p}_{z}}}{\hbar}u\right)  }}{\sqrt{2\pi\hbar}}du\right]  .
\label{wavtransf}%
\end{equation}

For $u$- dependent part of the spherical harmonics $Y_{lm}^{\prime}%
(u,\varphi)$ we get from (\ref{Yu}) and (\ref{wavtransf}),%
\begin{align}
Q_{lm}(p_{z})  &  \equiv N_{lm}\int_{-\infty}^{\infty}\frac{P_{l}^{m}(-\tanh
u)}{\cosh u}\frac{1}{\sqrt{2\pi\hbar}}{\exp\left(  i\frac{{{p}_{z}}}{\hbar
}u\right)  }du,\nonumber\\
&  =N_{lm}F(p_{z},\left[  \frac{P_{l}^{m}(-\tanh u)}{\cosh u}\right]  ),
\label{equ}%
\end{align}
where the Fourier transform $F(p,\left[  f(q)\right]  )$ of a function $f(q)$
is defined by,%
\begin{equation}
F(p,\left[  f(q)\right]  )\equiv%
%TCIMACRO{\dint }%
%BeginExpansion
{\displaystyle\int}
%EndExpansion
f(q)\frac{e^{ipq}}{\sqrt{2\pi}}dq. \label{FT}%
\end{equation}
For $\varphi$- dependent part of the spherical harmonics $Y_{lm}^{\prime
}(u,\varphi)$ we get in the $(p_{z},L_{z})$ representation a simple Kronecker
delta function $\delta_{L_{z},m\hbar}$ from (\ref{pzlz}). The original
spherical harmonics $Y_{lm}(\theta,\varphi)$ finally becomes $Y_{lm}%
^{\prime\prime}(p_{z},L_{z})$ in the ($p_{z},L_{z}$) representation,
\begin{equation}
Y_{lm}^{\prime}(u,\varphi)\rightarrow Y_{lm}^{\prime}(u,\varphi)\rightarrow
Y_{lm}^{\prime\prime}(p_{z},L_{z})\equiv Q_{lm}(p_{z})\delta_{L_{z},m\hbar}.
\end{equation}
The action of an operator $f(u,\varphi)$ on the wave function $\Psi
(u,\varphi)$ as $f(u,\varphi)\Psi(u,\varphi)$ in the $(p_{z},L_{z})$
representation is given by $f(-i\hbar\partial/\partial{{p}_{z}},L_{z})$ from
(\ref{wavtransf}),%
\begin{align}
f(u,\varphi)\Psi(u,\varphi)  &  \rightarrow\int_{0}^{2\pi}\frac{e^{-im\varphi
}}{\sqrt{2\pi}}d\varphi\left[  \int_{-\infty}^{\infty}(f(u,\varphi
)\Psi(u,\varphi))\frac{{\exp\left(  i\frac{{{p}_{z}}}{\hbar}u\right)  }}%
{\sqrt{2\pi\hbar}}du\right] \nonumber\\
&  =\int_{0}^{2\pi}\frac{e^{-im\varphi}}{\sqrt{2\pi}}d\varphi\left[
\int_{-\infty}^{\infty}\left(  f(-i\hbar\frac{\partial}{\partial{{p}_{z}}%
},\varphi)\frac{{\exp\left(  i\frac{{{p}_{z}}}{\hbar}u\right)  }}{\sqrt
{2\pi\hbar}}\right)  \Psi(u,\varphi)du\right] \nonumber\\
&  =f(-i\hbar\frac{\partial}{\partial{{p}_{z}}},L_{z})\int_{0}^{2\pi}%
\frac{e^{-im\varphi}}{\sqrt{2\pi}}d\varphi\left[  \int_{-\infty}^{\infty}%
\frac{{\exp\left(  i\frac{{{p}_{z}}}{\hbar}u\right)  }}{\sqrt{2\pi\hbar}}%
\Psi(u,\varphi)du\right] \nonumber\\
&  =f(-i\hbar\frac{\partial}{\partial{{p}_{z}}},L_{z})\Phi(p_{z},L_{z}).
\label{eigval}%
\end{align}
Here, same operator $f$ in different representations takes different variables
on which the operator depends differently.

Applying above results (\ref{wavtransf}), (\ref{equ}) and (\ref{eigval}) to
both sides of the eigenvalue function $L^{2}(u,\varphi)Y_{lm}^{\prime
}(u,\varphi)=l(l+1)\hbar^{2}Y_{lm}^{\prime}(u,\varphi)$, we have,%
\begin{equation}
L^{2}({{p}_{z}},L_{z})Q_{lm}(p_{z})\delta_{L_{z},m\hbar}=l(l+1)\hbar^{2}%
Q_{lm}(p_{z})\delta_{L_{z},m\hbar}.
\end{equation}
The $p_{z}$ dependent part $Q_{lm}(p_{z})$ satisfies following equation,
\begin{align}
&  N_{lm}\int_{0}^{2\pi}\left[  \left(  p_{z}^{2}+2i\hbar p_{z}\tanh
(u)+(m^{2}-1)\hbar^{2}\right)  \cosh^{2}(u)\frac{{\exp\left(  i\frac{{{p}_{z}%
}}{\hbar}u\right)  }}{\sqrt{2\pi\hbar}}\frac{P_{l}^{m}(-\tanh u)}{\cosh
u}du\right] \nonumber\\
&  =l(l+1)\hbar^{2}Q_{lm}(p_{z}). \label{main1}%
\end{align}
This equation (\ref{main1}) in fact has following two equivalent forms. One is
a differential equation from (\ref{eigval})
\begin{equation}
\left(  p_{z}^{2}+2i\hbar p_{z}\tanh(-i\hbar\frac{\partial}{\partial{{p}_{z}}%
})+(m^{2}-1)\hbar^{2}\right)  \cosh^{2}(-i\hbar\frac{\partial}{\partial
{{p}_{z}}})Q_{lm}(p_{z})=l(l+1)\hbar^{2}Q_{lm}(p_{z}). \label{main2}%
\end{equation}
Another is a difference equation with use of a relation: $\exp(\pm
au){\exp\left(  i{{p}_{z}}u/\hbar\right)  =\exp\left(  iu{({p}_{z}}\mp
ia\hbar)/\hbar\right)  }$,%
\begin{align}
l(l+1)\hbar^{2}Q_{lm}(p_{z})  &  =\frac{1}{2}\left[  p_{z}^{2}+(m^{2}%
-1)\hbar^{2}\right]  Q_{lm}(p_{z})\nonumber\\
&  +\frac{1}{4}\left[  p_{z}^{2}+(m^{2}-1)\hbar^{2}+2i\hbar p_{z}\right]
Q_{lm}(p_{z}-i2\hbar)\nonumber\\
&  +\frac{1}{4}\left[  p_{z}^{2}+(m^{2}-1)\hbar^{2}-2i\hbar p_{z}\right]
Q_{lm}(p_{z}+i2\hbar).
\end{align}
The similar difference equation appears in many systems, e.g. Morse oscillator
in momentum space \cite{morse}.

The following properties of $Q_{lm}(p_{z})$ are available. 1, Orthogonality
from Eq.(\ref{norm}):%
\begin{equation}
\int_{-\infty}^{\infty}Q_{l^{\prime}m^{\prime}}^{\ast}(p_{z})Q_{lm}%
(p_{z})dp_{z}=\delta_{l^{\prime}l}\delta_{m^{\prime}m}.
\end{equation}
2, Symmetries from Eq.(\ref{equ}):%
\begin{equation}
Q_{l(-m)}(p_{z})=(-1)^{m}Q_{lm}(p_{z}),Q_{lm}(-p_{z})=(-1)^{m}Q_{lm}(p_{z}).
\end{equation}
3, It can be verified that for a given quantum number $l$, they are $l+1$
linearly independent $l$th polynomials upon factors of sech$\left(  \pi
{{p}_{z}/(2\hbar)}\right)  $ corresponding to even $m=0,\pm2,\pm4,...$ \ or
csch$\left(  \pi{{p}_{z}/(2\hbar)}\right)  $ corresponding to odd $m=\pm
1,\pm3,\pm5,...$.

So far, in this section a dynamical $(p_{z},L_{z})$ representation on $S^{2}$
is established.

\section{Momentum spectrometer for some rotational states}

We now use the dynamical representation developed in section III to give the
momentum distribution of some rotational states, and then point out that this
distribution bears the feature of that for one-dimensional harmonic oscillator.

The first nine state functions of $Q_{lm}(p_{z})$ for $l=0$, $l=1$, and $l=2$
are from (\ref{equ}),%
\begin{equation}
Q_{0,0}(p_{z})=\frac{1}{2}\sqrt{\pi}\text{sech}\left(  \frac{\pi}{2}\frac
{{{p}_{z}}}{\hbar}\right)  , \label{Q0}%
\end{equation}%
\begin{equation}
Q_{1,0}(p_{z})=-\frac{1}{2}i\sqrt{3\pi}\frac{{{p}_{z}}}{\hbar}\text{sech}%
\left(  \frac{\pi}{2}\frac{{{p}_{z}}}{\hbar}\right)  ,\text{ }Q_{1,\pm1}%
(p_{z})=\pm\frac{1}{2}\sqrt{\frac{3\pi}{2}}\frac{{{p}_{z}}}{\hbar}%
\text{csch}\left(  \frac{\pi}{2}\frac{{{p}_{z}}}{\hbar}\right)  , \label{Q1}%
\end{equation}%
\begin{equation}
Q_{2,0}(p_{z})=-\frac{1}{8}\sqrt{5\pi}\left(  3(\frac{{{p}_{z}}}{\hbar}%
)^{2}-1\right)  \text{sech}\left(  \frac{\pi}{2}\frac{{{p}_{z}}}{\hbar
}\right)  ,\text{ }Q_{2,\pm1}(p_{z})=\pm\frac{1}{4}i\sqrt{\frac{15\pi}{2}%
}(\frac{{{p}_{z}}}{\hbar})^{2}\text{csch}\left(  \frac{\pi}{2}\frac{{{p}_{z}}%
}{\hbar}\right)  , \label{Q21}%
\end{equation}%
\begin{equation}
Q_{2,\pm2}(p_{z})=\frac{1}{8}\sqrt{\frac{15\pi}{2}}\left(  (\frac{{{p}_{z}}%
}{\hbar})^{2}+1\right)  \text{sech}\left(  \frac{\pi}{2}\frac{{{p}_{z}}}%
{\hbar}\right)  . \label{Q22}%
\end{equation}

For cases $l=0$, $l=3$, and $l=10$, the probability distributions $\left\vert
Q_{lm}(p_{z})\right\vert ^{2}$ of the dimensionless geometric momentum
$k_{z}\equiv p_{z}/\hbar$ for rotational states represented by spherical
harmonics $Y_{lm}(\theta,\varphi)$ are plotted in Figures 1, 2, and 3
respectively. In overall, they bear striking resemblance to the probability
amplitude of the dimensionless momentum for one-dimensional simple harmonic
oscillator. It is perfectly understandable that from the force operator
${\dot{p}}_{{i}}\equiv$ $[{p}_{{i}}{,H}]/(i\hbar)=-\{x_{i}/r,H\}$ $\sim-x_{i}$
with $\{U,V\}\equiv UV+VU$, we see that for the stationary state, the force is
restoring and proportional to the displacement.

\begin{figure}
[ptb]%
\includegraphics[scale=0.9]{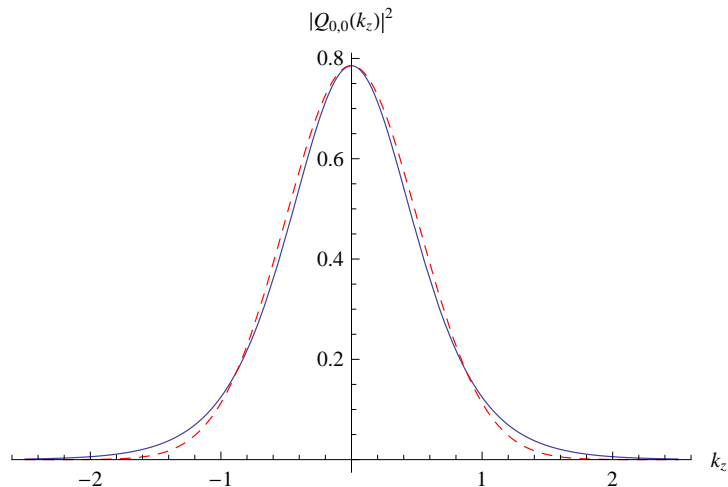}\caption{Geometric momentum distribution
density for the ground state of ground rotational state $Y_{0,0}=1/(\sqrt
{4\pi})$ (solid line), and the momentum distribution density for the ground
state of one-dimensional simple harmonic oscillator (dashed line). They are
almost identical. In all figures, the dimensionless momentum $k_{z} \equiv
p_{z}/\hbar$ is used. }\label{figure 1}
\end{figure}

\begin{figure}
[ptb]%
\includegraphics[scale=1.2]{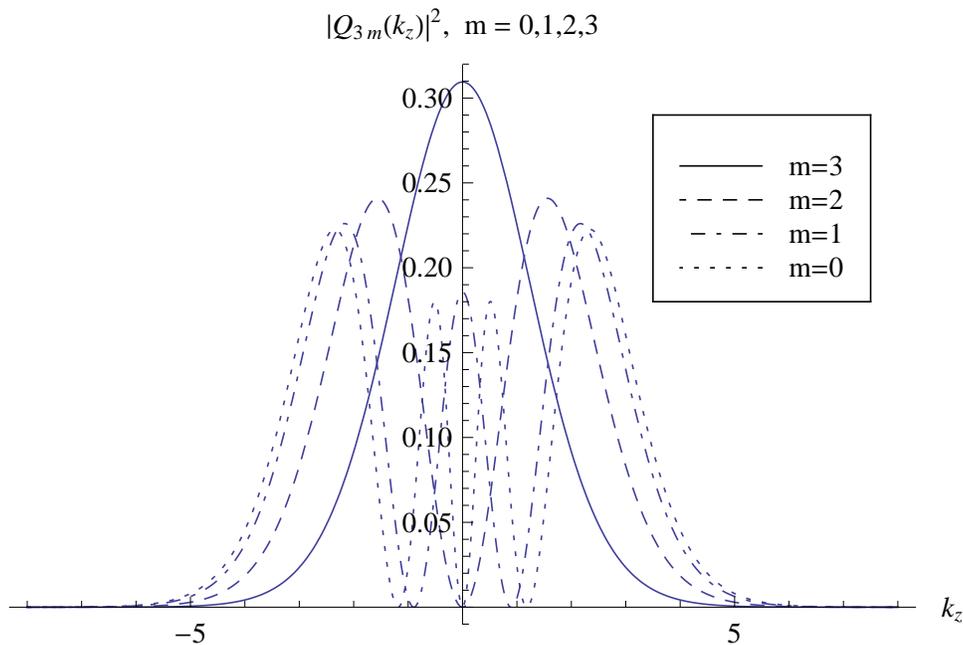}\caption{Geometric momentum distribution density for the rotational states $Y_{lm}(\theta,\varphi)$ with $l=1$ and
$m=0,1,2,3$, they have number of nodes $3,2,1,0$ respectively. It is worthy of stressing that for a given set ($l,m$),
i.e. each curve in this figure, behaves like a stationary harmonic oscillator state.}\label{figure 2}%

\end{figure}

\begin{figure}
[ptb]%
\includegraphics[scale=0.9]{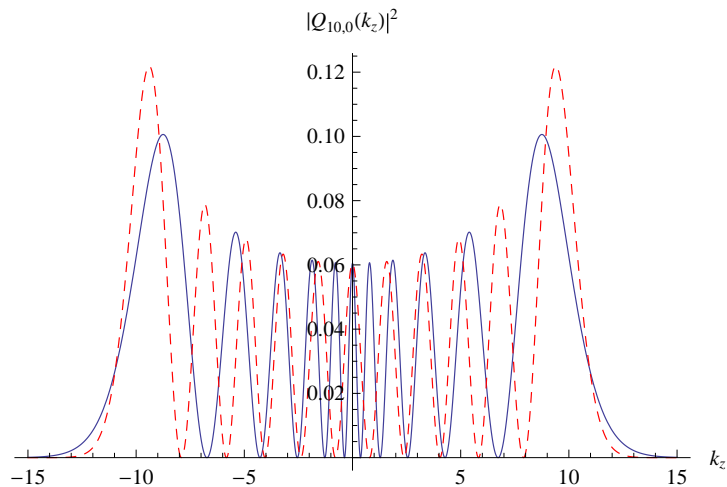}\caption{Geometric momentum distribution
density for rotational state $Y_{10,0}(\theta,\varphi)$ (solid line), and the
momentum distribution density for the $10$th excited state of one-dimensional
simple harmonic oscillator (dashed line). Since both probabilities in a small
interval $\Delta k_{z}$ are almost the same, they have the same classical
limit: the simple harmonic oscillation. }\label{figure 3}
\end{figure}

The rotation of a homonuclear diatomic molecule around its center of mass, and
the free rotation of spherical cage molecule $C_{60}$\ or $Au_{32}$
\cite{gong}\ around its center, etc.\cite{rotator5}, can be well modelled by a
spherical top. Here, there is a quantum uncertainty associated with degree of
freedom $r$, but we can choose \textit{a confining potential} $V(r)$\ such
that the uncertainty is minimum as $\Delta r\Delta p_{r}=\hbar/2$. Then it
raises a crucial problem: if we can subtract this radial momentum contribution
from the total one, whether our results $|Q_{lm}(p_{z})|^{2}$ can be
experimentally testable. Fortunately, the answer is affirmative, as shown below.

Resuming $p_{i}$ without multiplying the radius $r$ as doing in (\ref{hpx}%
)-(\ref{hpz}), and $[{x}_{{i}}{,p}_{j}]=i\hbar(\delta_{ij}-x_{i}x_{j}/r^{2}%
{)}$ \cite{liu11-1}. The ground state $Y_{00}(\theta,\varphi)=1/\sqrt{4\pi}$
is the minimum uncertainty state for three pairs of ($x_{i},p_{i}$) and
$\Delta x_{i}\Delta p_{i}=\hbar/3$. The state $Y_{00}(\theta,\varphi)$ bears
neither energy nor angular momentum, and the presence of zero-point the
momentum fluctuation $\Delta p_{i}=\hbar/(\sqrt{3}r)$ contradicts what
classical mechanics would indicate. With preparing these molecules into ground
state of rotation, the probability density of the geometric momentum
distribution is given by $|Q_{0,0}(p_{z})|^{2}$ (\ref{Q0}). With $r\approx
5.0$\AA \ for $C_{60}$, $\Delta p_{i}\approx0.07a.u.$($1a.u.=\hbar/a_{0}$,
with $a_{0}$ denoting the Bohr radius) and $r\approx1.0$\AA \ for $H_{2}$,
$\Delta p_{i}\approx0.3a.u.$ that seems within the resolution power of a
recently designed momentum spectrometer \cite{PSpectr1,PSpectr2,PSpectr3}.
Moreover, if it is possible to prepare these molecules into any excited
states, the momentum distributions $|Q_{lm}(p_{z})|^{2}$ given by Eq.
(\ref{equ}) have even high resolutions.

\section{Conclusions and Discussions}

How to understand quantum motions on a surface had been considered out of the
problem. This might be due to the fact that in elementary particle physics and
quantum gravity, physicists were acquainted with a fact that the outer space
of the universe had little effect on the inner one \cite{japan1993}. Thus,
consideration of the extrinsic curvature of two-dimensional surfaces was
thought sheer nonsense, and the intrinsic property of the surfaces suffices in
physics, which does not depend on whether they are embedded into the
three-dimensional Euclidean, even higher-dimensional, space or not
\cite{japan1993}. However, recent experiments demonstrate that the energy
spectrum on constrained motion on two-dimensional curved surface is
significantly influenced by the geometric potential depending on the extrinsic
curvature \cite{2010-2,epl}. We show that the geometric momentum is
indispensable to the geometric potential, and even more fundamental.

For motions on two dimensional spherical surface $S^{2}$, there is a new
dynamical symmetry obeying $SO(3,1)$ group whose six generators are the
Cartesian components of the geometric momentum $\mathbf{p}$ and the orbital
angular momentum $\mathbf{L}$, where the dependence of the geometric momentum
on extrinsic curvature, the mean curvature, reflects an embedding effect. From
the commutation relations $[{L}_{{i}}{,p}_{i}]=0$, $(i=1,2,3)$, we have three
complete sets of commuting observables, and they are equivalent with each
other upon a rotation of coordinates. Thus a novel dynamical representation
based on two observables, ($p_{z},L_{z}$) in the present paper, is
successfully constructed, and any states on $S^{2}$ can go through a momentum analysis.

Because the free rotation is ubiquitous in microscopic domain, we propose to
measure the momentum distribution of the state represented by spherical
harmonics to probe the embedding effect, once preparing the some molecules
into the state. This kind of experiments seems within reach of the present
nanotechnological capabilities
\cite{PSpectr1,PSpectr2,PSpectr3,rotator1,rotator2,rotator3,rotator4}.

\begin{acknowledgments}
This work is financially supported by National Natural Science Foundation of
China under Grant No. 11175063.
\end{acknowledgments}

\end{document}